\documentclass[final,5p,times,twocolumn,authoryear]{elsarticle}

\usepackage{amssymb}

\usepackage{color}

\journal{New Astronomy}

\begin{document}

\begin{frontmatter}



\title{Fractal structures for the Jacobi Hamiltonian of restricted three-body problem}


\author[utinam]{G. Rollin}
\ead{rollin@obs-besancon.fr}
\author[utinam]{J. Lages}
\ead{jose.lages@utinam.cnrs.fr}
\author[lpt]{D. L. Shepelyansky}
\ead{dima@irsamc.ups-tlse.fr}

\address[utinam]{Institut UTINAM, Observatoire des Sciences de l'Univers THETA, CNRS, Universit\'e de Franche-Comt\'e, 25030 Besançon,
France}
\address[lpt]{Laboratoire de Physique Th\'eorique du CNRS, IRSAMC, Universit\'e de Toulouse, UPS, 31062 Toulouse, France}

\begin{abstract}
We study the dynamical chaos and integrable motion 
in the planar circular restricted three-body problem
and determine the fractal dimension of the spiral strange repeller set of non-escaping orbits
at different values of mass ratio of binary bodies and of Jacobi integral of motion.
We find that the spiral fractal structure of the Poincar\'e section leads to
a spiral density distribution of particles remaining in the system.
We also show that the initial exponential drop of survival probability with time is followed 
by the algebraic decay related to the universal algebraic statistics of Poincar\'e
recurrences in generic symplectic maps.
\end{abstract}


\end{frontmatter}



\section{Introduction}
The restricted three-body problem was at the center of studies of dynamics in 
astronomy starting from the works of \citet{euler},
\citet{jacobi1836} and \citet{poincare}.
The progress in the understanding of this complex problem in XXth and XXIth centuries
is described in the fundamental books \citep{szebehely,henon1,henon2,valtonen06}.
As it was proven by \citet{poincare} in the general case this system is not integrable
and only the Jacobi integral is preserved by the dynamics \citep{jacobi1836}.
Thus a general type of orbits has a chaotic dynamics with a divided phase space
where islands of stability are embedded in a chaotic sea \citep{chirikov,lichtenberg92,ott}.

In this work we consider 
the Planar Circular Restricted Three-Body Problem (PCRTBP).
This is an example of a conservative Hamiltonian system 
(in a synodic or rotating reference frame of two binaries)
with two degrees of freedom. However, this is an open system 
since  some trajectories
can escape to infinity (be ionized) so that the general theories of 
leaking systems \citep{altmann} and naturally open systems \citep[e.g.][]{contopoulos04} find here their direct applications. 
It is known that such open systems are characterized by strange repellers
related to non-escaping orbits and by an exponential time decay of probability 
to stay inside the system. However, as we show, in the PCRTBP system with a divided
phase space one generally finds an algebraic decay of probability of stay
related to an algebraic statistics of Poincar\'e recurrences
in Hamiltonian systems 
\citep[see e.g.][and Refs. therein]{chirikov1981,karney,chirikov1984,meiss,chirikov1999,ketzmerick,shevchenko2010,frahm}. 
This effect appears due to long sticking of 
trajectories in a vicinity of stability islands 
and critical Kolmogorov-Arnold-Moser (KAM) curves. 
Thus an interplay of fractal structures and algebraic decay
in the PCRTBP deserves detailed studies.

Among the recent studies of the PCRTBP we point out 
the advanced results of \citet{nagler04,nagler05}
where the crash probability dependence on the size of large bodies has been studied
and the fractal structure of non-escaping orbits has been seen even if 
the fractal dimensions were not determined. 
This research line was extended in \citet{farrelly2004,farrelly2005} 
with a discussion of possible applications to the Kuiper-belt and 
analysis of various types of orbits in \citet{barrio,zotos}. 
The analysis of orbits in three dimensional case is reported in \citet{mako} and basin of escaping orbits around the Moon has been determined in \citet{assis14}.

In this work we determine the fractal dimension of non-escaping orbits 
for the PCRTBP with comparable masses of heavy bodies and consider the properties of
Poincar\'e recurrences and 
the decay probability of stay in this system.  The system description is given in Section 2,
the structure of strange repeller is analyzed in Section 3, 
the decay of Poincar\'e recurrences and 
probability of stay are studied in Section 4, a symplectic map description
of the dynamics is given in Section 5, discussion of the results is presented in Section 6.

\section{System description}

The PCRTBP system is composed of a test particle evolving
in the plane of a circular binary whose primaries have masses $m_1=1-\mu$ and $m_2=\mu$ 
with $m_1>m_2$. In the synodic frame the dynamics of the test particle 
is given by the Hamiltonian
\begin{equation}\label{eq:jacobihamiltonian}
H(p_x,p_y,x,y)=\displaystyle\frac{1}{2}\left(p_x^2+p_y^2\right)+yp_x-xp_y+V(x,y)
\end{equation}
where $x$ and $y$ are the test particle coordinates, $p_x=\dot x -y$ and $p_y=\dot y+x$ are
the corresponding canonically conjugated momenta, and
\begin{equation}\label{eq:potential}
V(x,y)=-\displaystyle\frac{\left(1-\mu\right)}{\left(\left(x-\mu\right)^2+y^2\right)^{1/2}}
-\displaystyle\frac{\mu}{\left(\left(x+\left(1-\mu\right)\right)^2+y^2\right)^{1/2}}
\end{equation}
is the gravitational potential of the two primaries.
Here the distance between primaries is $1$, the total mass $m_1+m_2=1$, 
the gravitational constant $\mathcal{G}=1$, 
consequently the rotation period of the binary is $2\pi$.
Hamiltonian (\ref{eq:jacobihamiltonian}) with potential (\ref{eq:potential}) 
represents the Jacobi integral of motion \citep{jacobi1836}. In the following we 
define the Jacobi constant as $C=-2H$. 
This Jacobi Hamiltonian describes also the planar dynamics 
of an electrically charged test particle experiencing a perpendicular magnetic field 
and a classical hydrogen-like atom with a Coulomb-like potential (\ref{eq:potential}).

\begin{figure}[!t]
\includegraphics[width=\columnwidth]{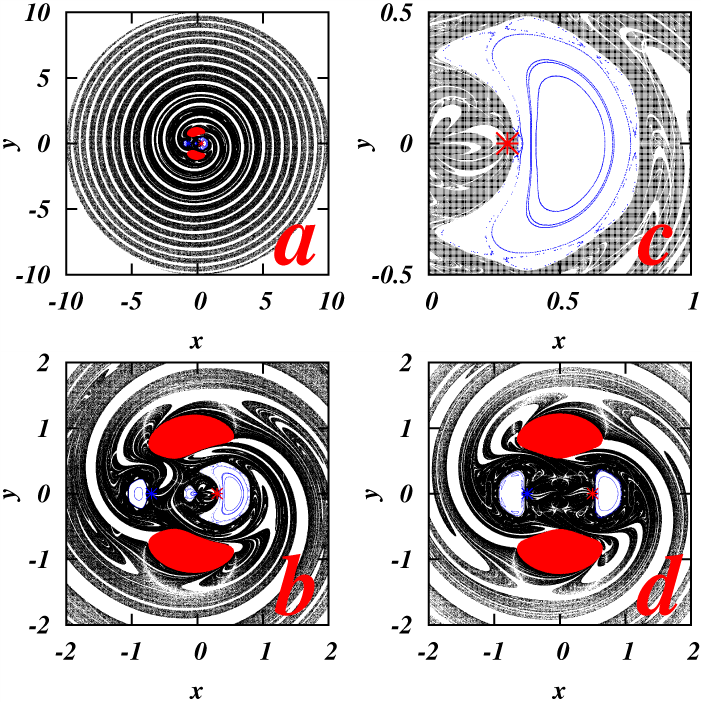}
\caption{$(x,y)$ - Poincar\'e sections of 
the Jacobi Hamiltonian (\ref{eq:jacobihamiltonian}) 
with $\dot r=0$ and $\dot\phi<0$.
Poincar\'e sections for a binary with $\mu=0.3$, $C=3$ are shown in panels: 
$(a)$ at a large scale, 
$(b)$ at an intermediate scale, $(c)$ close-up in the vicinity of the primary mass. 
Panel $(d)$ shows the Poincar\'e section for $\mu=0.5$ and $C=3$.
Red regions are forbidden since there $\dot x^2+\dot y^2<0$.
Black dots represent non-escaped orbits staying inside the $r<R_s=10$ region after time $t=10$.
Invariant KAM curves (blue dots) are obtained choosing initial conditions inside KAM islands.
The red (blue) star $\color{red}{\mathbf{*}}$ ($\color{blue}{\mathbf{*}}$) 
gives the position of the $1-\mu$ mass ($\mu$ mass). 
The Poincar\'e section is  obtained with orbits of $N=10^7$ test particles initially placed at random in the region $1.3\leq r\leq2.5$. Particles as escaped once $r>R_S$.}
\end{figure}

We aim to study the dynamics of particles evolving on escaping and non-escaping 
orbits around the binary.
We perform intensive numerical integration of the equations of motion derived 
from Hamiltonian (\ref{eq:jacobihamiltonian}) using an adaptive time step 
4th order Runge-Kutta algorithm with Levi-Civita regularization in 
the vicinity of the primaries \citep{levicivita20}.
The achieved accuracy is such as the integral of motion relative error is less than $10^{-9}$ ($10^{-5}$) for more than 91\% (99\%) of integration steps.
For different Jacobi constants $C$, 
we randomly inject up to $10^8$ test particles in the $1.3\leq r\leq 2.5$ ring with 
initial radial velocity $\dot r=0$ and initial angular velocity $\dot\phi<0$ 
($r$ and $\phi$ are polar coordinates in the synodic frame).
Each test particle trajectory is followed until the integration time attains 
$t_S=10^4$ or until the region $r>R_S=10$ is reached where we consider that 
test particles are escaped (ionized) from the binary.

\begin{figure}[!t]
\includegraphics[width=\columnwidth]{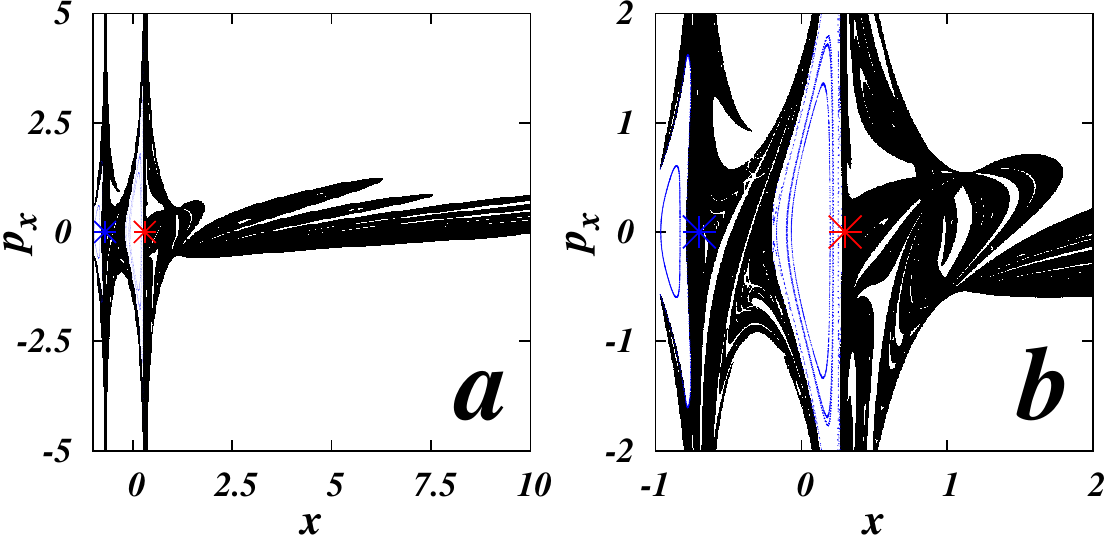}
\caption{$(p_x,x)$ - Poincar\'e section of the Jacobi Hamiltonian (\ref{eq:jacobihamiltonian}) 
with $y=0$ and $p_y>0$ for a binary with $\mu=0.3$, $C=3$ (corresponding to Fig.~1$a,b,c$). 
Panel $(a)$: Poincar\'e section at large scale; panel $(b)$: zoom in the vicinity of primaries.
Black dots represent non-escaped orbits staying inside the $r<R_s=10$ region after time $t=10$.
Blue dots represent bounded orbits inside stability islands.
The red (blue) star $\color{red}{\mathbf{*}}$ ($\color{blue}{\mathbf{*}}$) 
gives the position of the primary (secondary) mass as in Fig.~1.
The Poincar\'e section is obtained with the same orbits as in Fig.~1.}
\end{figure}

\section{Strange repeller structures}

In phase space, orbits are embedded in a three-dimensional surface defined by 
the Jacobi constant $C$. In order to monitor particle trajectories we choose 
a two-dimensional surface defined by an additional condition. Here we choose either 
the condition $(\dot r=0,\dot \phi<0)$ to represent Poincar\'e section as a $(x,y)$-plane 
(Figs.~1,5,6,7 and 10) or the condition $(y=0,p_y>0)$ to 
represent Poincar\'e section as a $(p_x,x)$-plane (Fig.~2).
A similar approach was also used in \citet{nagler04,nagler05}.

\begin{figure}[!t]
\includegraphics[width=\columnwidth]{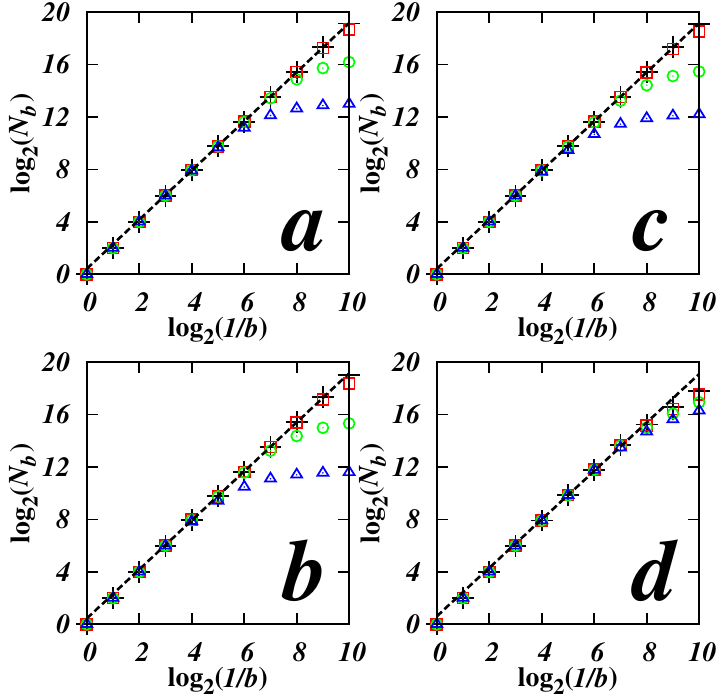}
\caption{Number of boxes $N_b$ covering at scale $b$
non-escaped orbits structure (strange repeller) 
appearing in $(x,y)$ - Poincar\'e section of 
the Jacobi Hamiltonian (\ref{eq:jacobihamiltonian}).
Box-counting computation is performed in an annulus square consisting in a square region $-3.9\leq x,y\leq3.9$ deprived of its central square region $-1.3\leq x,y\leq1.3$.
The annulus square is divided into 8 equal square areas of linear size $dl_0=1.3$. We average the box counting $N_b$ over the 8 squares. At scale $b$ each square is divided into $1/b^2$ boxes of linear size $dl=b\,dl_0$.
Box-counting results are shown for Poincar\'e section of
orbits staying in the $r<R_S=10$ 
disk after $t=3$ (black crosses), $t=10$ (red squares), $t=30$ (green circles), 
$t=50$ (blue triangles).
The fractal dimension $D$ of the strange repeller 
is the slope of the affine function $\log_2 N_b=f(\log_2(1/b))$.
Keeping orbits staying in the $r<R_S$ disk after time $t=10$ ($t=3$) 
we obtain a strange repeller fractal dimension
$(a)$ $D=1.8711 \pm 0.0100$ ($D=1.8732 \pm 0.0105$) for $\mu=0.3$ and $C=3$ (see Fig.~1$a,b,c$),
$(b)$ $D=1.8657 \pm 0.0117$ ($D=1.8690 \pm 0.0129$) for $\mu=0.4$ and $C=3$ (see Fig.~10$b$),
$(c)$ $D=1.8700 \pm 0.0077$ ($D=1.8722 \pm 0.0084$) for $\mu=0.3$ and $C=2.6$ (see Fig.~5$a$),
$(d)$ $D=1.8349 \pm 0.0484$ ($D=1.8464 \pm 0.0436$) for $\mu=0.3$ and $C=3.4$ (see Fig.~5$d$).
Fits have been performed in the scale range $2^{4}\leq 1/b\leq 2^{8}$. 
We used $N=10^8$ $(a,b,c)$, $N=10^6$ $(d)$ test particles initially distributed at random in the $1.3\leq r\leq2.5$ ring.
The fractal dimension has been computed with
$(a)$
$N_{t>3}=39526570$,
$N_{t>10}=9933333$,
$N_{t>30}=768282$,
$N_{t>50}=83290$ points,
$(b)$
$N_{t>3}=26743797$,
$N_{t>10}=6550163$,
$N_{t>30}=372871$,
$N_{t>50}=25037$ points,
$(c)$
$N_{t>3}=37610948$,
$N_{t>10}=8721338$,
$N_{t>30}=419296$,
$N_{t>50}=39891$ points,
$(d)$
$N_{t>3}=8569720$,
$N_{t>10}=5447406$,
$N_{t>30}=2245927$,
$N_{t>50}=1083887$ points.
The curves bend down for the smallest scales $b$ when $1/b^2$ becomes of the order of the number of collected points in Poincar\'e section.
}
\end{figure}

We show in Fig.~1 (panels $a,b,c$) an example of $(x,y)$ - Poincar\'e section of 
the Jacobi Hamiltonian (\ref{eq:jacobihamiltonian}) for mass parameter $\mu=0.3$ and 
Jacobi constant $C=3$. Red regions correspond to forbidden zones 
where particles would have imaginary velocities.
Inside central islands in the close vicinity of primaries blue points mark out 
regular and chaotic orbits of bounded motion. In particular, 
the KAM invariant curves \citep{lichtenberg92} 
can be seen \textit{e.g.} in Fig.~1$c$. 
In Fig.~1$a$, the trace of non-escaped chaotic orbits (black points) remaining inside 
the disk $r<R_S=10$  after time $t=10$ defines a set of points forming two spiral 
arms centered on the binary center of mass. This set has a 
spiral structure of strange repeller 
since orbits in its close vicinity rapidly move away from 
the set and consequently from the binary. 
The fractal property results in a self-similar structure clearly 
seen by zooming to smaller and smaller scales 
(see Fig.~1$a,b,c$). 
This fractal structure remains stable in respect to moderate variation of mass ratio
$\mu$ as it is illustrated in Fig.~1 (panels $b,d$).
A strange repeller structure is also clearly visible 
in the corresponding Poincar\'e section in $(p_x,x)$ plane shown in 
Fig.~2 for $\mu=0.3$, $C=3$. 

\begin{figure}[!t]
\includegraphics[width=\columnwidth]{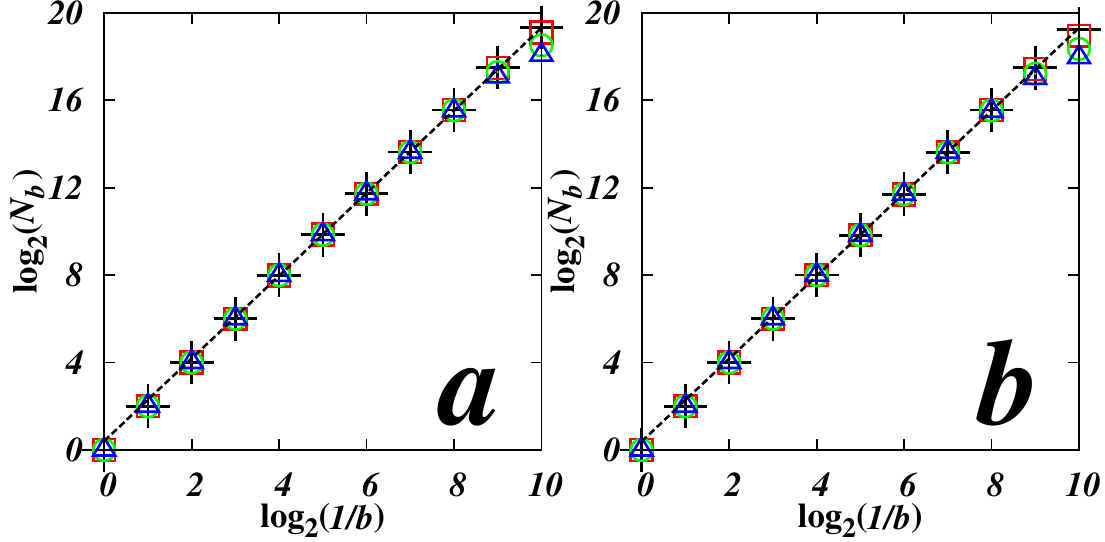}
\caption{Number of boxes $N_b$ covering at scale $b$
non-escaped orbits structure (strange repeller) 
appearing in $(x,y)$ - Poincar\'e section of 
the Jacobi Hamiltonian (\ref{eq:jacobihamiltonian}).
Box-counting computation is performed as in Fig.~3.
Box-counting results are shown for Poincar\'e section of
orbits staying in the $r<R_S=100$ 
disk after $t=3$ (black crosses), $t=10$ (red squares), $t=30$ (green circles), 
$t=50$ (blue triangles).
Keeping orbits staying in the $r<R_S=100$ disk after time $t=3,10,30,50$ 
we obtain a strange repeller fractal dimension
$(a)$ $D=
1.8908\pm0.00876,
1.8900\pm0.00858,
1.8874\pm0.00754,
1.8799\pm0.00480
$ for $\mu=0.3$ and $C=3$ (see Fig.~1$a,b,c$),
$(b)$ $D=
1.8916\pm0.0129,
1.8911\pm0.0127,
1.8869\pm0.0110,
1.8786\pm0.0086
$ for $\mu=0.4$ and $C=3$ (see Fig.~10$b$).
Fits have been performed in the scale range $2^{4}\leq 1/b\leq 2^{8}$. 
We used $N=10^8$ test particles initially distributed at random in the $1.3\leq r\leq2.5$ ring.
The fractal dimension has been computed with
$(a)$
$N_{t>3}=38090345$,
$N_{t>10}=18470667$,
$N_{t>30}=7588914$,
$N_{t>50}=4574705$
points,
$(b)$
$N_{t>3}=27206778$,
$N_{t>10}=14259496$,
$N_{t>30}=6542192$,
$N_{t>50}=4286763$
points.
}
\end{figure}

\begin{figure}[!t]
\includegraphics[width=\columnwidth]{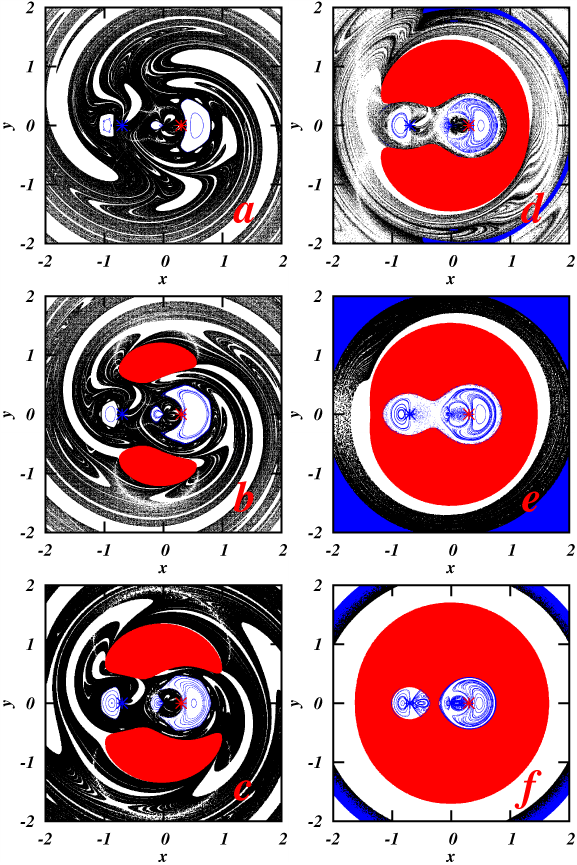}
\caption{$(x,y)$ - Poincar\'e sections of the Jacobi Hamiltonian (\ref{eq:jacobihamiltonian}) 
with $\dot r=0$ and $\dot\phi<0$
for $\mu=0.3$ and Jacobi constant $(a)$ $C=2.6$, $(b)$ $C=3$, $(c)$ $C=3.2$, 
$(d)$ $C=3.4$, $(e)$ $C=3.6$, and $(f)$ $C=4$.
Red regions are forbidden since there $\dot x^2+\dot y^2<0$.
Black dots represent non-escaped orbits staying inside the $r<R_s=10$ region after time $t=10$.
Blue dots represent bounded orbits.
The red (blue) star $\color{red}{\mathbf{*}}$ ($\color{blue}{\mathbf{*}}$) gives 
the position of the primary (secondary) mass as in Fig.~1.
Poincar\'e sections have been obtained analyzing orbits of $N=10^7$ $(a,b,c)$ and 
$N=10^5$ $(d,e,f)$ particles initially placed in the $1.3\leq r\leq2.5$ region. Particles are considered as escaped once $r>R_S$.}
\end{figure}

The fractal dimension of the strange repeller
is determined using box-counting method \citep{lichtenberg92,ott} 
as $D=\lim_{b\rightarrow0}\ln{N_b}/\ln{(1/b)}$ where $N_b$ is at scale $b$ the number 
of at least once visited boxes in the Poincar\'e section. 
The box-counting fractal dimension of the strange repeller 
presented in Fig.~1$a,b,c$ is $D \approx 1.87$ (Fig.~3$a$). 
This fractal dimension value is computed 
from the strange repeller structure formed by orbits staying 
in the disk $r<R_S=10$  after time $t=10$. 
We see from Fig.~3 that the fractal dimension remains practically the same for the parameters considered here $\mu=0.3,0.4$ and $C=2.6,3,3.4$.
When the escape radius is increased
up to $R_S=100$ (Fig.~4) the fractal dimension value is the same as for an escape radius $R_S=10$ (Fig.~3).
Thus the obtained value of $D$ is not affected by the escape cut-off distance $R_S$.
Also, as seen in Fig.~4, the fractal dimension remains practically 
the same if we consider strange repeller structures obtained after time 
$t=3$, $10$, $30$ and $50$.
Hence even for short times the strange repeller structure 
is already well defined and perdures for greater times since $D$ 
is constant (at least here up to $t=50$). 
Throughout this work, for the sake of clarity we choose to present Poincar\'e sections 
for orbits staying in the $r<R_S=10$ disk after time $t=10$.

\begin{figure}[!t]
\includegraphics[width=\columnwidth]{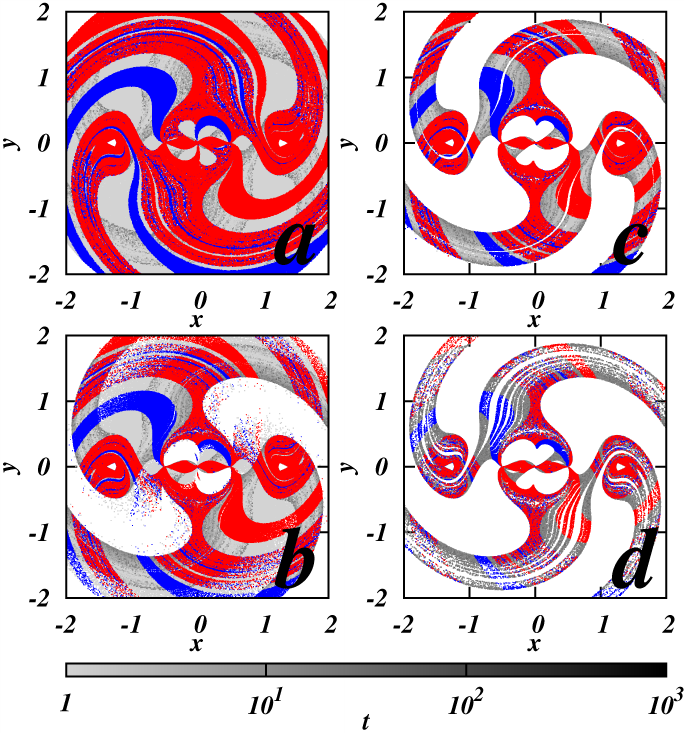}
\caption{$(x,y)$ - Poincar\'e sections of the Jacobi Hamiltonian 
(\ref{eq:jacobihamiltonian}) with $\dot r=0$ and $\dot\phi<0$
for primary bodies with radius $r_b=0.01$ and for $\mu = 0.5$ and $C=1$.
The panels show traces of orbits associated with particles escaping at time 
$(a)$ $t_{esc}>0$ (\textit{i.e.} all computed orbits are shown),
$(b)$ $t_{esc}>0.01$,
$(c)$ $t_{esc}>1$,
$(d)$ $t_{esc}>10$.
The gray scale bar shows the time when particles pass through the Poincar\'e section.
Light gray (dark gray) points have been obtained at $t \approx 1$ ($t \approx 10^3$).
Red (blue) points have been obtained from orbits crashing on the $\mu$ (1-$\mu$) primary mass.
Initially $10^6$ particles have been 
randomly distributed in the $1.3 \leq r \leq 2.5$ ring.
Fig.~5 (4th panel for $r_b=0.01$) in \citet{nagler04} is similar to panel $(a)$.
}
\end{figure}

\begin{figure}[!t]
\includegraphics[width=\columnwidth]{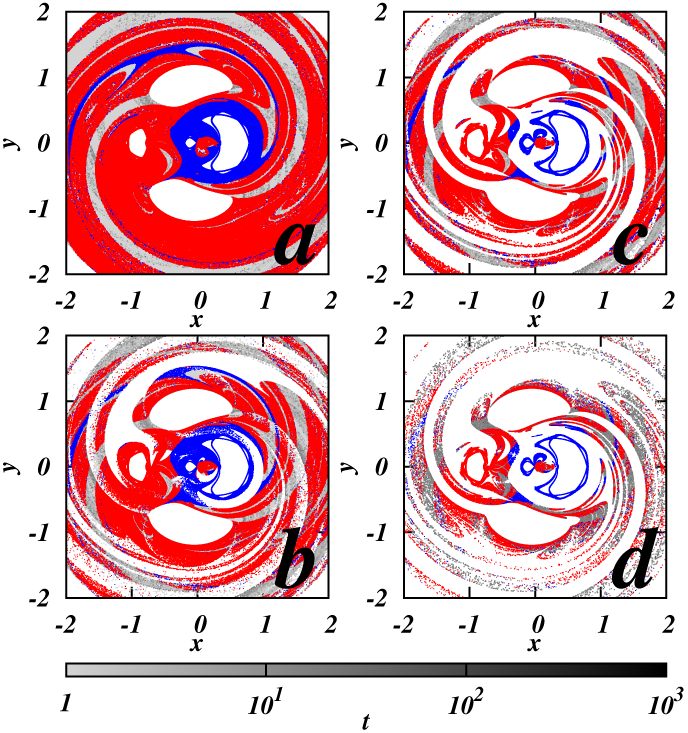}
\caption{$(x,y)$ - Poincar\'e sections of the Jacobi Hamiltonian 
(\ref{eq:jacobihamiltonian}) with $\dot r=0$ and $\dot\phi<0$ 
for primary bodies with radius $r_b=0.01$ and for $\mu = 0.3$ and $C=3$.
The panels show
$(a)$ all points after $t = 0$,
$(b)$ all points after $t = 0.01$,
$(c)$ all points after $t = 1$,
$(d)$ all points after $t = 10$.
The gray scale bar shows the time when particles pass through the Poincar\'e section.
Light gray (dark gray) points have been obtained at $t \approx 1$ ($t \approx 10^3$).
Red (blue) points have been obtained from orbits crashing on the $\mu$ (1-$\mu$) primary mass.
Initially $10^5$ particles have been 
randomly distributed in the $1.3 \leq r \leq 2.5$ ring.
}
\end{figure}

Fig.~5 shows $(x,y)$ - Poincar\'e sections for the mass parameter $\mu=0.3$ and 
for different Jacobi constants from $C=2.6$ to $C=4$. The strange repeller structure constituted 
by non-escaping orbits is progressively expelled from the primaries vicinity as $C$ increases. 
At $C=2.6$, $3$, $3.2$, $3.4$ (Fig.~5$a,b,c,d$) 
non-escaping trajectories may still pass close by each one of the primaries. 
The strange repeller still dominates the phase portrait with a fractal dimension 
decreasing down to $D \approx 1.84$ for $C=3.4$ (Fig.~3$d$). For greater values $C=3.6$, $4$ (Fig.~5$e,f$)  
the forbidden zone insulates the immediate vicinity of the primaries from trajectories 
coming from regions beyond $r\sim 1$. Regular and chaotic trajectories corresponding 
to particles gravitating one or the two primaries are confined 
in the very central region \citep{valtonen06}. 
The strange repeller is confined in a narrow ring located in the region $r\sim 1.5$. 
Beyond that region we observe nearly stable circular orbits (blue dots) corresponding 
to particles gravitating the whole binary with a radius $r \sim 2$. For these orbits stable 
means that these orbits have not escaped from the disk $r<R_S$ 
during the whole integration duration $t_S$.

Unless otherwise stated, we have deliberately omitted the class of orbits crashing primaries. According to 
\citet{nagler04,nagler05} the crash basin scales as a power law $r_b^\alpha$ where $r_b$ is the radius 
of the primary mass and $\alpha\sim0.5$. In this work we choose a radius of 
$r_b=10^{-5}$ for the two primaries which gives two percent of crashing orbits and 
an area of about one percent for crash basin not visible in the presented Poincar\'e sections.
The sets of non-escaping orbits shown in Fig.~1 are also distinguishable in the studies 
\citep{nagler04,nagler05} devoted to crashing orbits but not studied in details. 
For example Fig.~1$d$ presents a Poincar\'e section for the Copenhagen problem case 
($\mu=0.5$) with $C=3$ which is similar to the Poincar\'e section presented in the study 
\citep{nagler04} Fig.~3 right column middle row for $C=2.85$.

The time evolution of density of non-escaped particles is shown in Fig.~6 for $\mu=0.5$
and Fig.~7 for $\mu =0.3$ for the case of primary bodies of relatively
large radius $r_b=0.01$ (such a size is also available
in Fig.~5 in \citet{nagler04}). These data clearly show that the strange repeller structure
is established on rather short time scales with $t \sim 1$. 
We also see that for such a value of $r_b$ the measure of crashed orbits
gives a visible contribution to the measure of non-escaping orbits of strange repeller.

Finally we note that in our computations we determined the fractal dimension $D$ of
trajectories non-escaping in future \citep[see a similar situation considered for the Chirikov
standard map with absorption in][]{ermann}. According to the general relations
known for the strange fractal sets in dynamical systems
the fractal dimension $D_0$ of the invariant repeller set 
(orbits non-escaping neither in the future 
nor in the past) satisfies the relation $D_0=2 (D-1)$  \citep{ott}.
Thus for the typical value we have in Fig.~1 at $\mu=0.3$ with $D\approx1.87$ we obtain
$D_0\approx1.74$.

\begin{figure}[!t]
\includegraphics[width=\columnwidth]{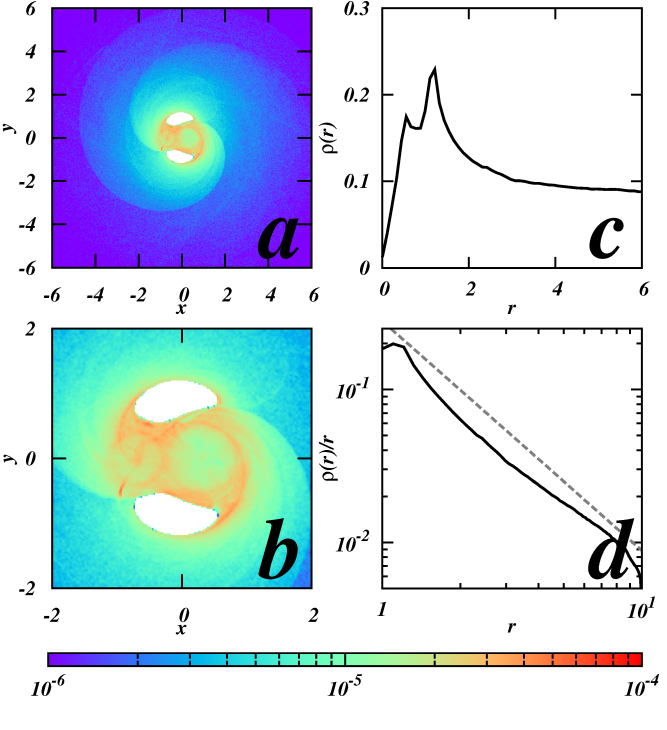}
\caption{$(a)$ Snapshot of the remaining particles at $t=10$ for $\mu=0.3$ and $C=3$. 
The number of particles initially injected in the $1.3\leq r\leq2.5$ ring is $N=10^8$,
at time $t=10$ there are $N_t=13302225$ particles remaining inside the circle
$r\leq R_S= 10$, colors show the surface density of particles $\rho_s$ in the plane $(x,y)$,
color bar gives the logarithmic color scale of density with levels corresponding to a proportion of 
the $N_t$ remaining particles;
(b) same as panel (a) but on a smaller scale;
(c) linear density $\rho(r)=dN_t/dr$;
(d) angle averaged surface density $\rho(r)/r = 2\pi <\rho_s(r)>$,
the dashed line shows the slope $\propto 1/r^{3/2}$. 
}
\end{figure}

We can expect that the spiral fractal structure, clearly present in the plane $(x,y)$
of the Poincar\'e sections (see e.g. Fig.~1),
will give somewhat similar traces for the surface (or area) density $\rho_s = dN_t/dxdy$
of particles $N_t$ remaining in the system at an instant moment of time $t$.
A typical example of surface density, corresponding to Fig.~1$a,b,c$,
is shown in Fig.~8. Indeed, we find a clear spiral structure of $\rho_s(x,y)$
similar to the spiral structure of the strange repeller of Fig.~1$a,b,c$.
Of course, here in Fig.~8 we have the projected density
of particles in $(x,y)$ plane taken at all values of $\dot r$ 
(and not only at $\dot r =0$ as in Fig.~1), This leads to
a smoothing of the fractal structure of the Poincar\'e section
but the spiral distribution of density $\rho_s$ remains well visible.
We also note that the angle averaged density
of remaining particles $<\rho_s(r)> \propto 1/r^{3/2}$
has a radial dependence on $r$ being similar
to those found for the dark matter density obtained 
in the symplectic simulations of scattering and capture of 
dark matter particles on binary system \citep[see e.g. Fig.~4$a$ in][]{lagesdmp,rollin2}.

\begin{figure}[!t]
\includegraphics[width=\columnwidth]{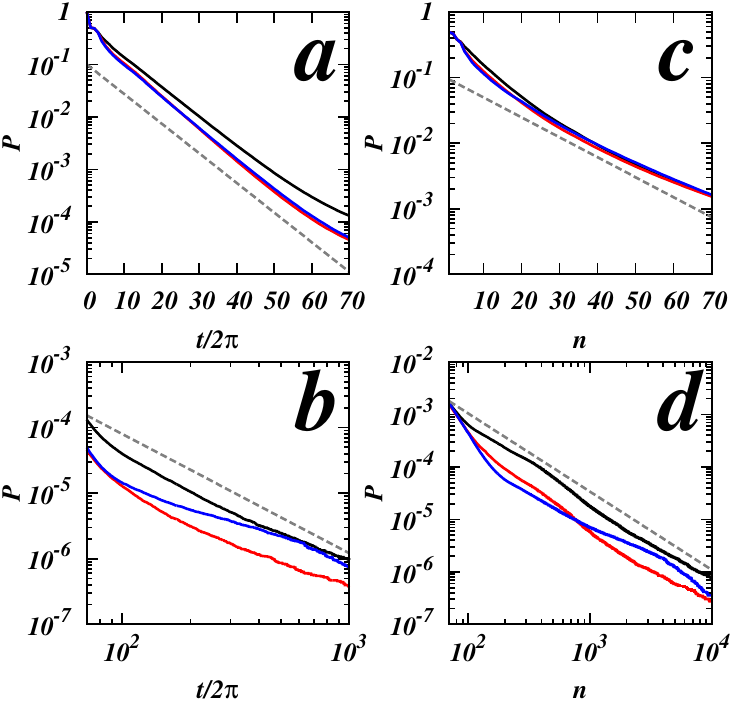}
\caption{Survival probability $P$ of particles as a function of time $t$ (left panels, 
binary period is $2\pi$) and as a function of the number $n$ 
of successive Poincar\'e section crossings 
(right panels) for $C=3$ and binaries with mass parameter $\mu=0.3$ (black curve), 
$\mu=0.4$ (red curve), $\mu=0.5$  (blue curve). Survival probabilities are shown 
in semi-log scale (top panels) and in log-log scale (bottom panels).
Dashed lines show
$(a)$ exponential decay $P \propto \exp(-t/\tau_s)$ with $1/\tau_s=0.13$,
$(b)$ algebraic decay $P(t) \propto 1/t^\beta$ 
with the Poincar\'e exponent $\beta=1.82$,
$(c)$ exponential decay $P(n) \propto \exp(-n/\tau_s)$ with $1/\tau_s=0.07$,
$(d)$ algebraic decay $P(n) \propto 1/n^\beta$ with $\beta=1.49$. Initially $10^8$ particles have been 
randomly distributed in the $1.3 \leq r \leq 2.5$ ring. Escape radius is $R_S=10$.
}
\end{figure}

\section{Poincar\'e recurrences and probability decay}

We determine numerically the probability $P(t)$ to stay inside the system
for time larger than $t$. For that $N$ particles are randomly distributed in the range
$1.3 \leq r \leq 2.5$ and then the survival probability $P(t)$ 
is defined as the ratio  $P(t)=N_t/N$, where
$N_t$ is the number of particles remained inside the system 
with $r<R_S=10$ at times larger than $t$. This survival probability
is proportional to the integrated probability of Poincar\'e recurrences
\citep[see e.g.][]{chirikov1999,frahm}.

The typical examples of the decay $P(t)$ are shown in Fig.~9.
At relatively small time scales with $t<100$ the decay can be approximately 
described by an exponential decay with a decay time $\tau_s \sim 10$.
Indeed, for the strange dynamical sets (e.g. strange attractors) one obtains usually
an exponential decay since there are no specific sticking regions in such strange sets
\citep{chirikov1984}.

However, at larger time scales $t >100$ we see the appearance of the algebraic 
decay of probability corresponding to the algebraic statistics of Poincar\'e
recurrences discussed for symplectic maps   
\citep[see e.g.][]{chirikov1999,ketzmerick,frahm}.
When the decay time is measured in number of crossings of the Poincar\'e section $n$ we 
obtain the 
Poincar\'e exponent $\beta$ of this decay $\beta =1.49$ being close to the values
$\beta \approx 1.5$ found in the symplectic maps. However, if the time 
is measured in number of rotations of binaries $t/2\pi$ then we find a somewhat
large value of $\beta$ (see Fig.~9). We explain this deviation a bit later.

\begin{figure}[!b]
\includegraphics[width=\columnwidth]{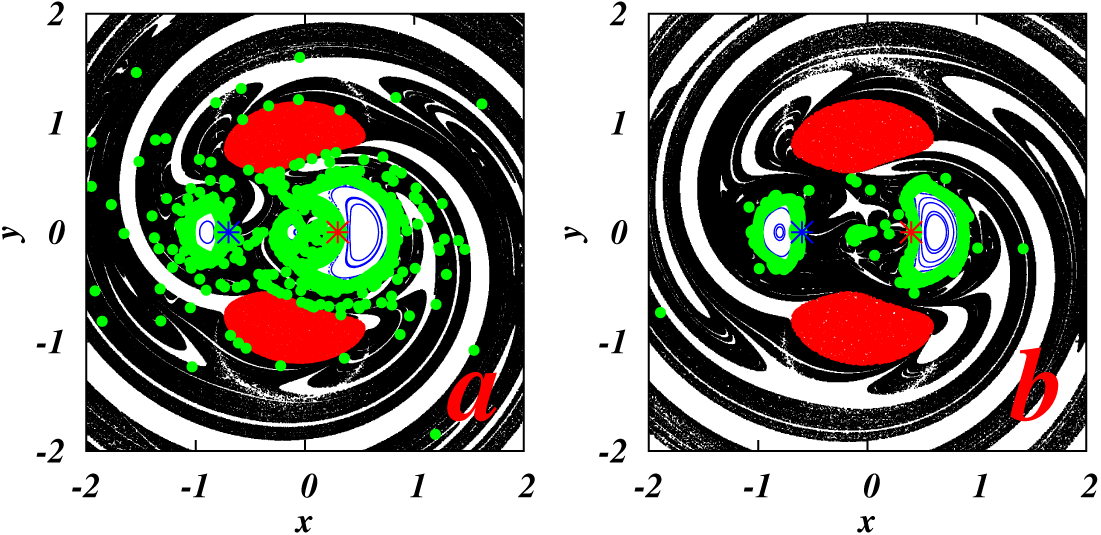}
\caption{$(x,y)$ - Poincar\'e sections of the Jacobi Hamiltonian (\ref{eq:jacobihamiltonian}) 
with $\dot r=0$ and $\dot\phi<0$ for $C=3$
in the case of $(a)$ a $\mu=0.3$ binary and $(b)$ a $\mu=0.4$ binary.
Red regions are forbidden since there $\dot x^2+\dot y^2<0$.
Black dots represent non-escaped orbits staying inside the $r<R_S=10$ region after time $t=10$.
Blue dots represent bounded orbits on KAM curves inside integrable islands.
Green plain circles mark out non-escaped orbits remaining inside 
the disk $r<R_s$  after time $t=500$.
The red (blue) star $\color{red}{\mathbf{*}}$ ($\color{blue}{\mathbf{*}}$) gives 
the position of the primary (secondary) mass. Each of Poincar\'e sections is obtained 
from orbits of $N=10^7$ particles initially placed in the $1.3\leq r\leq2.5$ region; 
these particles are considered as escaped once $r>R_s$.}
\end{figure}

The properties of orbits surviving in the system for long times are
shown in Fig.~10. We see that such orbits are concentrated in the vicinity of critical KAM curves
which separate the orbits of strange repeller from 
the integrable islands with KAM curves.  This is exactly the situation discussed in
the symplectic maps \citep[see e.g.][]{chirikov1999,ketzmerick,frahm}.
Thus the asymptotic decay of survival probability is determined by long time sticking orbits
in the vicinity of critical KAM curves. The detail analytical explanation of this
generic phenomenon is still under debates \citep[see e.g.][]{meiss2015}.

\begin{figure}[!t]
\includegraphics[width=\columnwidth]{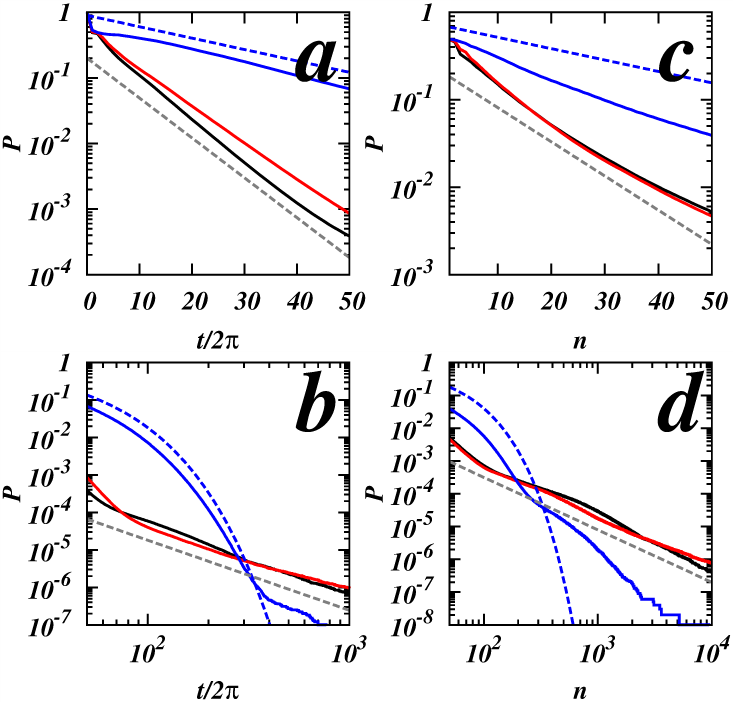}
\caption{Survival probability $P$ of particles as a function of time $t$ (left panels, 
binary period is $2\pi$) and as a function of the number $n$ 
of successive Poincar\'e section crossings 
(right panels) for $\mu =0.3$ and 
$C=2.6$ (black curve), $C=3$ (red curve), $C=3.4$  (blue curve). 
Survival probabilities are shown in lin-log scale (top panels) 
and in log-log scale (bottom panels).
Blue dashed lines show exponential decay $P(t) \propto \exp(-t/\tau_s)$ with $1/\tau_s=0.04$ (a,b)
and $P(n) \propto \exp(-n/\tau_s)$ with $1/\tau_s=0.03$ (c,d).
Gray dashed lines show
$(a)$ exponential decay $P(t)$ with $1/\tau_s=0.14$,
$(b)$ algebraic decay $P(t)$ with $\beta=1.87$,
$(c)$ exponential decay $P(n)$ with $1/\tau_s=0.09$,
$(d)$ algebraic decay $P(n)$ with $\beta=1.6$.
Initially $10^8$ particles have been randomly distributed in the $1.3 \leq r \leq 2.5$ ring.
Escape radius is $R_S=10$.
}
\end{figure}

At fixed $\mu=0.3$ the variation of decay properties of $P(t)$ with the Jacobi constant $C$
is shown in Fig.~11.
For $C=3.4$ the exponential decay of $P(t)$ is much slower than for the cases $C=3.6$ and $C=3$ (there is a factor $3.5$ between corresponding characteristic time scales $\tau_s$). We attribute this behavior to the fact that the forbidden zone encloses almost completely binary components letting for particles only a small route to binary components in comparison with cases $C=2.6$ and $C=3$ for which the route to gravitational perturbers is not constrained. Meanwhile, particles have a limited access to the chaotic component (see Fig.~5$d$ where it forms a narrow peanut shell around central stability islands) so that the sticking in the vicinity of critical KAM curves is reduced emphasizing the exponential decay. For $C=2.6,3$ at long time scales (Fig.~11d) we have a well visible algebraic decay $P(n)$ with $\beta \sim 1.5$ (we may assume that increasing sufficiently the number of initially injected particle for the case $C=3.4$, blue line in Fig.~11d would also exhibit at large time scales the same $\beta \sim 1.5$ algebraic decay). However, algebraic decay of $P(t)$ for $C=2.6,3$ at large time scale (Fig.~11b) still have somewhat different value of $\beta$.

\begin{figure}[!t]
\includegraphics[width=\columnwidth]{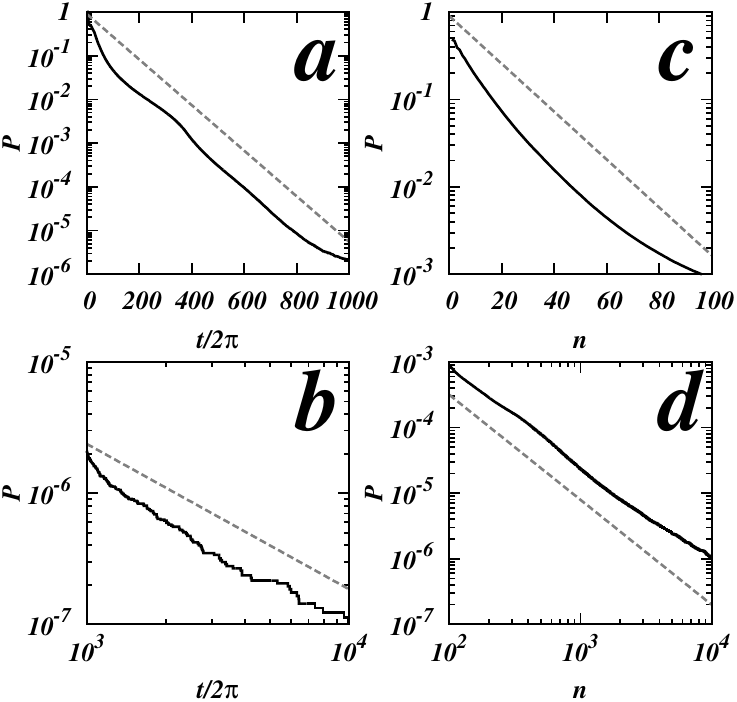}
\caption{Survival probability $P$ of particles as a function of time $t$ (left panels, 
binary period is $2\pi$) and as a function of the number $n$ 
of successive Poincar\'e section crossings 
(right panels) for $\mu=0.3$, $C=3$ in the case of an escape radius $R_S=100$.
Dashed curves show 
$(a)$ exponential decay $P(t)$ with $1/\tau_s=0.012$,
$(b)$ algebraic decay $P(t)$ with $\beta=1.11$,
$(c)$ exponential decay $P(n)$ with $1/\tau_s=0.062$,
$(d)$ algebraic decay $P(n)$ with $\beta=1.40$.
Initially $10^8$ particles have been 
randomly distributed in the $1.3 \leq r \leq 2.5$ ring.
}
\end{figure}

The origin of this difference for $P(t)$ becomes clear from Fig.~12
where we show the data similar to those of Figs.~9,11 at $\mu=0.3$
but with the escape radius $R_S=100$. We see that the decay properties of $P(n)$
remain practically unchanged that confirms the generic features of obtained results
for $\beta$ (indeed, stability islands do not affect dynamics at $r \sim R_S=100$).
However, the value of $\beta$ for $P(t)$ is significantly reduced to $\beta \approx 1.1$.
We explain this by the fact that in the usual time units the measure of chaotic component 
at large distances becomes dominant and the escape time is determined
simply by a Kepler rotation period which becomes larger for large $r$ values. This leads to the decay exponent $\beta =2/3$ for $P(t)$ as
discussed in \citet{borgonovi} and explains the variation of $\beta$ with $R_S$.
However, when the time is counted in the number of orbital periods
the divergence of the orbital period at large $r$ values (or small coupling energies)
does not affect the decay and we obtain the Poincar\'e exponent $\beta \approx 1.5$
being independent of $R_S$.

We note that the recent studies of survival probability decay
in the PCRTBP also report the value of $\beta \approx 1.5$ \citep{kovacs}.

\begin{figure}[!t]
\includegraphics[width=\columnwidth]{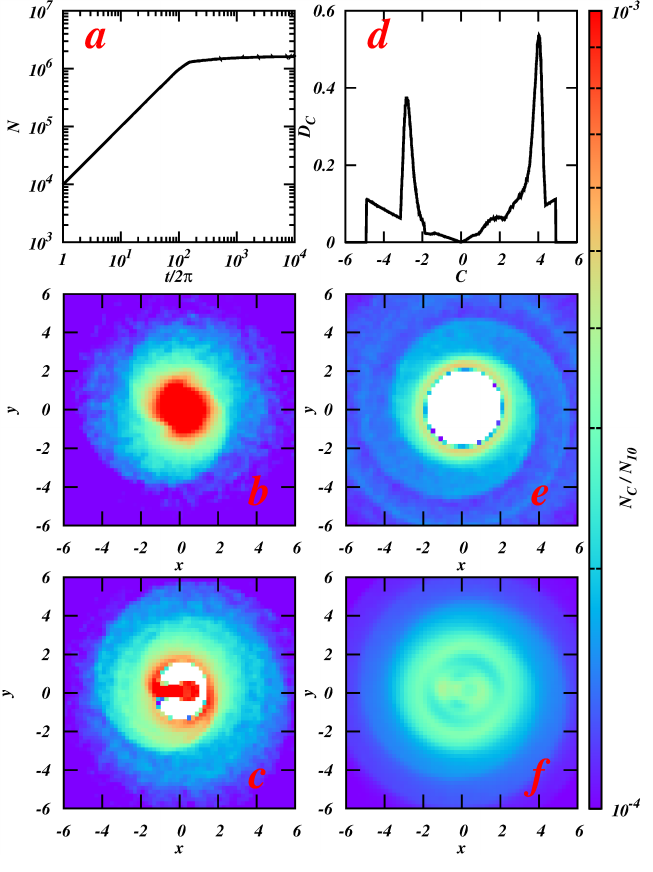}
\caption{Density of particles in the steady state.
(a) Number $N$ of particles in the $r<100$ region of the binary as a function of time $t$. We have continuously injected $N=10^8$ particles placed randomly on $r=100$ circle with close to parabolic but hyperbolic trajectories in the 2-body paradigm. Steady state is achieved around $t\sim t_0=10^3$.
The surface density of particles $\rho_s(x,y)$ is shown for Jacobi constants $C=2.6$ (b), $C=3.4$ (c) and $C=4$ (e). Each surface density has been obtained from $1000$ surface density snapshots taken at regular time interval after $t_0=10^3$ once steady state is attained.
Color bar gives the logarithmic color scale of density with levels corresponding to the ratio of $N_C/N_{10}$ where $N_C$ is the number of particles in a given cell of size $0.1\times0.1$ and $N_{10}$ is the number of particles in the $r<10$ region. Panel (d) shows the normalized steady state distribution of Jacobi constants of particles present in the $r<100$ region. Panel (f) shows the steady state surface density build with all particles in $r<100$ region.
}
\end{figure}

\section{Symplectic map description}

Finally we discuss the case when particles in the sidereal reference 
frame scatter on the binary with relatively large values of perihelion distance
$q \gg 1$. Such a case corresponds to large $|C| \gg 1$.
For $q \gg 1$ the dynamics of particle in the field of two binaries is
approximately described by the symplectic map of the form
${\bar w} =  w + F(x); {\bar x} = x + {\bar w}^{-3/2}$ where $w=-2E$
is the particle energy, $x$ is the phase of binary rotation (in units of $2\pi$) 
when the particle is located at its perihelion and
$F(x) \propto \mu $ is a periodic function of $x$ 
\citep{petrosky,halley,shevchenkohal,rollin1,rollin2};
bars above $w$ and $x$ mark new values after one rotation around the binary.
The amplitude $F_{max}$ decreases exponentially with increasing $q$.
Usually, one considers the case of $\mu \ll 1$ (e.g. Sun and Jupiter)
but our studies show that this form remains valid even for $\mu \sim 0.5$
if $q \gg 1$. At $\mu \ll 1$ we have  $F_{max} \ll 1$ 
and the escape time $t_i$ becomes very large
$t_i \propto 1/{F_{max}}^2$ being much larger than the 
Lyapunov time scale. In this situation the fractal dimensions $D$ and $D_0$ are
very close to $D=D_0=2$ \citep{ott,ermann} and the fractal effects
practically disappear. Due to that this case is not interesting for
the fractal analysis.

\section{Discussion}

We analyzed the PCRTBP dynamical system and showed that
for moderate mass ratio of primary bodies $\mu \sim 0.5$ the 
Poincar\'e section is characterized by a strange repeller with the fractal dimension 
having typical values $D \approx 1.87 $ $(D_0 \approx 1.74)$
at moderate values of the Jacobi constant $C \sim 2$.
At the same time certain islands of integrable motion are still present.
Such a structure of the Poincar\'e section leads to an exponential decay
of survival probability in the system on short time scales
followed by the algebraic decay with the Poincar\'e exponent $\beta \approx 1.5$
being similar to the values known for the statistics of Poincar\'e recurrences in
generic symplectic maps. For the small mass ratio $\mu \ll 1$ the escape times
becomes very large and the fractal dimension becomes close to the usual value $D=2$.

It is interesting to note that the strange repeller structure (see Figs.~1,5,6,7,10)
reminds the structure of spiral galaxies 
\citep[see e.g.][]{milkyway}.
In fact the fractal features of galaxies have been studied extensively
by various groups \citep{elmegreen,alfaro1,alfaro2}
which obtained from observation data  the fractal dimensions for the plane density being around
$D_g \approx 1.3 - 1.7$ for different galaxies.
Some of these values \citep[e.g.][with $D_g \approx 1.7$]{alfaro1} are similar to those
obtained here for the Poincar\'e section of PCRTBP.
Our results show that the spiral fractal structure of
the Poincar\'e section of PCRTBP leads to a spiral structure
of global density of particles $\rho_s$ remaining in the system
(see Fig.~8).
Thus we make a conjecture that the spiral
structure of certain galaxies can be linked to the
underlying spiral fractals appearing in the dynamics of 
particles in binary systems.
Of course, such a conjecture requires more detailed
analysis beyond the present paper such as taking into account the third dimension.
Also whether or not the spiral structure remains in the steady state is an important question. To bring some elements of response to the latter question we have injected at $r=100$ particles with close-to-parabolic hyperbolic trajectories in 2-body paradigm. We use a Maxwellian distribution for initial velocities $f(v)dv=\sqrt{54/\pi}\left(v^2/u^3\right)\exp\left(-3v^2/2u^2\right)dv$ with $u=0.035$ and a homogeneous distribution for perihelion $q\in[0;3]$ parameters. We checked that our results in steady state are independent of initial distributions. We present on Fig.~13 the surface density of particles $\rho_s(x,y)$ in the steady state attained around $t_0\sim10^3$ (Fig.~13a). We clearly see that the two arms spiral structure remains in the steady state, we show examples for particles with trajectories characterized by (b) $C=2.6$, (c) $C=3.4$, (e) $C=4$ (example not shown for $C=3$ is analogous to Fig~8a). However when all injected particles are considered independently of the Jacobi constant $C$ we obtain a blurred spiral (Fig.~13f). The contributions of different Jacobi constants are consequently mixed according to the distribution Fig.~13d. We remark that the results presented in Fig.~13 share similarities with those obtained in the study of chaotic trajectories in spiral galaxies \citep{harsoula11,contopoulos12,contopoulos13}.
On the basis of obtained results
we make a conjecture about existence of certain 
links between observed fractal dimensions
of galaxies \citep{elmegreen,alfaro1,alfaro2} 
and fractal spiral repeller structure studied here.
We think that the further studies of fractal structures in binary systems
will bring new interesting results.



\bibliographystyle{elsarticle-harv}
\bibliography{bibbinary}





\end{document}